\documentstyle[12pt]{article}

\begin{document}

\smallskip

\begin{flushright}
{\it PP-IOP-99/20 \\}
{\it GUTPA/00/04/01 \\}
\end{flushright}
\vskip 1in

\begin{center}
{\Large {\bf Lepton Number Violation in Supersymmetric Grand Unified
Theories }} \vspace{30pt}

{\bf J.L. Chkareuli }$^{a}${\bf \ and C.D. Froggatt }$^{b}$

\medskip

$^{a}$ {\em \ }{\small {\it Institute of Physics, Georgian Academy of
Sciences, 380077 Tbilisi, Georgia}}

$^{b}$ {\small {\it Department of Physics and Astronomy, Glasgow University,
Glasgow G12 8QQ, Scotland}}

\vskip .5in {\bf Abstract}
\end{center}

{\ {We argue that the nature of the global conservation laws in
Supersymmetric Grand Unified Theories is determined by the basic vacuum
configuration in the model rather than its Lagrangian. It is shown that the
suppression of baryon number violation in a general ($R$-parity violating)
superpotential can naturally appear in some extended $SU(N)$ SUSY GUTs
which, among other degenerate symmetry-breaking vacua, have a missing VEV
vacuum configuration giving a solution to the doublet-triplet splitting
problem. We construct $SU(7)$ and $SU(8)$ GUTs where the effective lepton
number violating couplings immediately evolve, while the baryon number
non-conserving ones are safely projected out as the GUT symmetry breaks down
to that of the MSSM.However at the next stage, when SUSY breaks, the
radiative corrections shift the missing VEV components to some nonzero
values of order $M_{SUSY}$, thereby inducing the ordinary Higgs doublet
mass, on the one hand, and tiny baryon number violation, on the other. So, a
missing VEV solution to the gauge hierarchy problem leads at the same time
to a similar hierarchy of baryon vs lepton number violation. }}

\thispagestyle{empty}\newpage

\section{Introduction}

The Standard Model (SM), while being extremely successful in describing
interactions of quarks and leptons at low energies, still has many
unanswered questions. Among these one problem is predominant: the
unification of all elementary forces within the framework of a simple gauge
theory and the ensuing hierarchy of mass scales in particle physics at large
and small distances. Possible solutions to this problem are commonly related
to supersymmetry (SUSY) \cite{nilles} and Grand Unified Theories (GUT) \cite
{2}. At present they receive some indirect (largely qualitative)
experimental support from the apparent lightness of the Higgs boson, the
values of gauge couplings given by precision measurements and the heavy top
quark mass. At low energies the SUSY GUT turns into the Minimal
Supersymmetric Standard Model (MSSM). Of course the MSSM can certainly be
considered on its own as a simple SUSY extension of the Standard Model,
leaving aside for the moment the question of unification.

However, no matter which level of theory is considered, there is one point
that crucially distinguishes SUSY from non-SUSY models. This is that they do
not contain the automatic accidental symmetries, corresponding to baryon (B)
and lepton (L) number conservation, which are present in the ordinary
Standard Model. Does this mean that B and L number can be violated in the
SUSY context or must some special protecting symmetry be postulated in the
MSSM? Usually, the requirement of B and L number conservation in the MSSM is
indeed satisfied by postulating the existence of some multiplicative
discrete symmetry called R-parity \cite{nilles}. An exact R-parity ($RP$)
implies that SUSY particles should be produced in pairs and that the
lightest SUSY particle (LSP) is stable.

On the other hand, there is no fundamental reason to prefer models with
exact $RP$ over those with broken $RP$ in the framework of the
supersymmetric SM, where not only fermions but also their scalar
superpartners automatically become the carriers of lepton and baryon
numbers. Thereby, among the basic renormalisable couplings in the
low--energy MSSM superpotential,one would generally expect to find the
lepton and baryon number violating ones~
\begin{equation}
\Delta W=\mu _{i}L_{i}H_{u}+\lambda _{ijk}L_{i}L_{j}\overline{E}_{k}+\lambda
_{ijk}^{\prime }L_{i}Q_{j}\overline{D}_{k}+\lambda _{ijk}^{\prime \prime }%
\overline{U}_{i}\overline{D}_{j}\overline{D}_{k}.  \label{1}
\end{equation}
Here, $i$, $j$, $k$ are generation indices and a summation is implied
(colour and weak isospin indices are suppressed); $L_{i}$($Q_{j}$) denote
the lepton (quark) $SU(2)$--doublet superfields and $\overline{E}_{i}$($%
\overline{U}_{i}$, $\overline{D}_{i}$) are $SU(2)$--singlet lepton ($up$%
--quark, $down$--quark) superfields; $\mu _{i}$ are mass parameters which
mix lepton superfields with the $up$--type Higgs superfield $H_{u}$, while 
$%
\lambda _{ijk}$ ($\lambda _{ijk}=-\lambda _{jik}$), $\lambda _{ijk}^{\prime
} $ and $\lambda _{ijk}^{\prime \prime }$ ($\lambda _{ijk}^{\prime \prime
}=-\lambda _{ikj}^{\prime \prime }$) are dimensionless couplings. The first
three terms in (\ref{1}) violate lepton number, while the last violates
baryon number.

While SUSY-inspired B number violation (BNV) leads in general to
unacceptably fast proton decay and must be highly suppressed, SUSY-inspired
L number violation (LNV) could readily occur at a level consistent with
present experimental constraints, but large enough for the observation of
some of its spectacular manifestations \cite{lnv} at present or future
colliders. Remarkably enough, instead of $RP$, another (gauge) discrete
symmetry could appear in the MSSM: superstring-inherited $Z_{3}$ baryon
parity \cite{Z3}, which strongly protects B number and allows for L number
violation only. Thus, as long as it is not in conflict with any
phenomenology, SUSY-inspired lepton number violation merits further detailed
investigation, both theoretically and experimentally.

{}From the theoretical point of view, the principal question concerns the
search for a Grand Unified framework within which, while treating quarks and
leptons equally, L-violation should be allowed at the same time as
B-conservation. Unfortunately, the discrete symmetries acceptably protecting
B-conservation while allowing L-violation in the MSSM, such as the above
mentioned $Z_{3}$ baryon parity, transform quarks and leptons differently
and, thereby, are incompatible with the known GUTs. Nevertheless several
extended GUT models have been constructed \cite{lnvGUTs}, where the
coexistence of lepton number violation and baryon number conservation can in
principle be arranged. This is typically achieved by introducing at the
Planck scale high-dimensional operators, involving Higgs and matter
multiplets in some exotic representations of the underlying GUT symmetry,
and then imposing additional custodial symmetries to ensure that only the
required set of LNV high-order operators is allowed. These operators become
the renormalisable LNV couplings (\ref{1}) at lower energies after the GUT
symmetry breaks at the GUT scale $M_{GUT}$.

Despite some progress, one has the uneasy feeling that such a solution to
this problem looks rather artificial, as it is generically correlated
neither with the nature of the GUT nor with its breaking pattern. Instead,
we suggest that it is just the breaking pattern of the underlying GUT
symmetry which could give a fundamental reason for the difference in
treatment of the baryon and lepton numbers of the matter particles involved
in GUTs. We show that a suppression of baryon number violating interactions
in the superpotential (\ref{1}) naturally occurs in some $SU(N)$ SUSY GUTs
where a missing VEV vacuum configuration develops, which also gives a
solution to the doublet-triplet splitting problem \cite{jgk}. We construct
explicit examples of $RP$-violating $SU(7)$ and $SU(8)$ GUTs where the
effective LNV couplings immediately evolve from the GUT scale, while the
baryon number non-conserving ones are safely projected out by the missing
VEV vacuum configuration breaking the GUT symmetry down to that of the MSSM.
However, at the next stage when SUSY breaks, radiative corrections shift the
missing VEV to some nonzero value of order $M_{SUSY} $ and induce BNV
violating couplings with hierarchically small coupling constants $\lambda
_{ijk}^{\prime \prime }=O(M_{SUSY}/M_{GUT})$, which appear to be of
phenomenological interest \cite{lnv}.

\section{Missing VEV solutions in $SU(N)$ GUTs}

The most elegant solution to the gauge hierarchy problem in supersymmetric 
$%
SU(N) $ GUTs could well be related to the existence of a missing VEV vacuum
configuration \cite{dim}, according to which the basic adjoint scalar $%
\Sigma _{j}^{i}$ ($i,j=1,...,N$) does not develop a VEV in some of the
directions in $SU(N)$ space. Through its coupling with a pair of Higgs
fields $H$ and $\overline{H}$, their masses are split in a hierarchical way
so as to have light weak doublets breaking electroweak symmetry and giving
masses to up and down quarks, on the one hand, and superheavy colour
triplets mediating proton decay, on the other. However, it is well known
\cite{dim} that a missing VEV solution can not appear in $SU(N)$ GUTs in the
ordinary one-adjoint scalar case. This is due to the presence of a cubic
term $\Sigma ^{3}$ in the general Higgs superpotential $W$ leading to the
unrealistic trace condition $Tr\Sigma ^{2}=0$ for the missing VEV vacuum
configuration, unless there is a special fine-tuned cancellation between $%
Tr\Sigma ^{2}$ and driving terms stemming from other parts of the
superpotential $W$.

\subsection{The two-adjoint alternative}

So, it seems the only way to obtain a natural missing VEV solution in 
$SU(N)$
theories is to exclude the cubic term $\Sigma ^{3}$ from the superpotential,
by imposing some extra reflection symmetry on the adjoint supermultiplet $%
\Sigma $
\begin{equation}
\Sigma \to -\Sigma  \label{1a}
\end{equation}
On its own the elimination of the $\Sigma ^{3}$ term leads to the trivial
unbroken symmetry case. However the inclusion of higher even-order $\Sigma $
terms (supposedly inherited from superstrings or induced by gravitational
corrections) in the effective superpotential leads to an all-order missing
VEV solution, as was shown in recent papers \cite{jgk}. Alternatively one
can introduce another adjoint scalar $\Omega $ with only renormalisable
couplings appearing in $W$.

Let us consider briefly the high-order term case first. The $SU(N)$
invariant superpotential for an adjoint scalar field $\Sigma $ conditioned
also by the gauge $Z_{2}$ reflection symmetry (\ref{1a})

\begin{equation}
W_{A}=\frac{1}{2}m\Sigma ^{2}+\frac{\lambda _{1}}{4M_{P}}\Sigma ^{4}+\frac{%
\lambda _{2}}{4M_{P}}\Sigma ^{2}\Sigma ^{2}+...  \label{1b}
\end{equation}
contains, in general, all possible even-order $\Sigma $ terms scaled by
inverse powers of the (conventionally reduced) Planck mass $M_{P}=(8\pi
G_{N})^{-1/2}\simeq 2.4\cdot 10^{18}$ GeV. It is readily shown that the
necessary condition for any missing VEV solution to appear in the $%
SU(N)\otimes $ $Z_{2}$ invariant superpotential $W_{A}$ is the tracelessness
of all the odd-order $\Sigma $ terms

\begin{equation}
Tr\Sigma ^{2s+1}=0\mbox{ , } \quad s=0,1,2,...  \label{1c}
\end{equation}
This condition uniquely leads to a missing VEV pattern of the type

\begin{eqnarray}
&&~~~~~~~~\hspace{0.5cm}~~~~\quad N-k\mbox{ \quad \quad }k/2%
\mbox{ \qquad
\quad }k/2  \nonumber \\
&<&\Sigma >=\sigma Diag(\overbrace{~0~...~0~}~,\overbrace{~1~...~1~}~,%
\overbrace{~-1~...-1~})\mbox{,}  \label{1d}
\end{eqnarray}
where the VEV value $\sigma $ is calculated using the $\Sigma $ polynomial
taken in $W_{A}$ (\ref{1b}). The vacuum configuration (\ref{1d}) gives rise
to a particular breaking channel of the $SU(N)$ GUT symmetry

\begin{equation}
SU(N)\rightarrow SU(N-k)\otimes SU(k/2)\otimes SU(k/2)\otimes
U(I)_{1}\otimes U(I)_{2}\mbox{ ,}  \label{1e}
\end{equation}
which we will discuss in some detail a little later. So we conclude from
Eqs. (\ref{1d}, \ref{1e}) that a missing VEV solution could actually exist,
with the ordinary MSSM gauge symmetry $SU(3)_{C}\otimes SU(2)_{W} \otimes
U(1)_{Y}$ surviving at low energies, provided that $N \ge 7$.

The superpotential (\ref{1b}) can be viewed as an effective one, following
from an ordinary renormalisable two-adjoint superpotential with the second
heavy adjoint scalar integrated out. Hereafter, although both approaches are
closely related, we deal for simplicity with the two-adjoint case. So let us
consider a general $SU(N)$ invariant renormalisable superpotential for two
adjoint scalars $\Sigma $ and $\Omega$, also satisfying the gauge-type $%
Z_{2} $ reflection symmetry ($\Sigma \to -\Sigma $, $\Omega \to \Omega $)
inherited from superstrings:
\begin{equation}
W_{A}=\frac{1}{2}m\Sigma ^{2}+\frac{1}{2}M_{P}\Omega ^{2}+\frac{1}{2}h\Sigma
^{2}\Omega +\frac{1}{3}\lambda \Omega ^{3}.  \label{2}
\end{equation}
Here the second adjoint $\Omega $ can be considered as a state originating
from a massive string mode with the Planck mass $M_{P}$. The basic adjoint 
$%
\Sigma $ may be taken at another well motivated scale $m\sim
M_{P}^{2/3}M_{SUSY}^{1/3}\sim O(10^{13})$ GeV \cite{yan} where, according to
many string models, the adjoint moduli states $(1_{c},1_{w})$, $%
(1_{c},3_{w}) $ and $(8_{c},1_{w})$ (in a self-evident $SU(3)_{C}\otimes
SU(2)_{W}$ notation) appear. In the present context these states can be
identified as just the non-Goldstone remnants $\Sigma _{0,}$ $\Sigma _{3}$
and $\Sigma _{8} $ of the relatively light adjoint $\Sigma $ which breaks $%
SU(N)$ in some way. However, all our conclusions remain valid for any
reasonable value of $m $, which is the only mass parameter (apart from $%
M_{P} $) in the model considered.

As a general analysis of the superpotential $W_{A}$ (\ref{2}) shows \cite
{jgk}, there are just four possible VEV patterns for the adjoint scalars $%
\Sigma $ and $\Omega $: (i) the trivial unbroken symmetry case, $\Sigma
=\Omega =0$; (ii) the single-adjoint condensation, $\Sigma =0$, $\Omega \neq
0$; (iii) the $^{\prime \prime }$parallel$^{\prime \prime }$ vacuum
configurations, $\Sigma \propto \Omega $ and (iv) the $^{\prime \prime }$%
orthogonal$^{\prime \prime }$ vacuum configurations, $Tr(\Sigma \Omega )=0$.
The Planck-mass mode $\Omega $, having a cubic term in $W_{A}$, in all
non-trivial cases develops a standard $^{\prime \prime }$single-breaking$%
^{\prime \prime }$ VEV pattern

\begin{eqnarray}
&&~~~~~~~~~~\hspace{0.5cm}~\qquad N-k~~~~~\hspace{0.5cm}~\qquad ~k  
\nonumber
\\
&<&\Omega >\mbox{ }=\omega Diag(\overbrace{~1~...~1~}~,~\overbrace{- \frac{%
N-k}{k}...-\frac{N-k}{k}})\mbox{,}  \label{3a}
\end{eqnarray}
which breaks the $SU(N)$ GUT symmetry to

\begin{equation}
SU(N)\rightarrow SU(N-k)\otimes SU(k)\otimes U(I)\mbox{ .}  \label{3b}
\end{equation}
However, in case (iv), the basic adjoint $\Sigma $ develops the radically
new missing VEV vacuum configuration (\ref{1d}), thus giving a $^{\prime
\prime }$double breaking$^{\prime \prime }$ of $SU(N)$ to (\ref{1e}). Using
the approximation $\frac{h}{\lambda }>>\frac{m}{M_{P}}$, which is satisfied
for any reasonable values of the couplings $h$ and $\lambda $ in the generic
superpotential $W_{A}$ (\ref{2}), the VEV values are given by

\begin{equation}
\omega =\frac{k}{N-k}\frac{m}{h}\quad \mbox{, }\quad \sigma =\left( 
\frac{2N%
}{N-k}\right) ^{1/2}\sqrt{mM_{P}}/h  \label{vev}
\end{equation}
respectively. Surprisingly, just the light adjoint $\Sigma $ develops the
largest VEV in the model which, for a properly chosen adjoint mass $m$ and
coupling constant $h$, can easily come up to the string scale $M_{str}$ (see
\cite{su7tex}).

Furthermore, as concluded above, one must consider $SU(N)$ GUTs with $N \ge
7 $, in order to have the standard gauge symmetry $SU(3)_{C}\otimes
SU(2)_{W}\otimes U(1)_{Y}$ remaining after the breaking (\ref{1e}). As is
easily seen from Eqs. (\ref{1d}, \ref{1e}), there are two principal
possibilities: the weak-component and colour-component missing VEV solutions
respectively. If it is granted that the ''missing VEV subgroup'' $SU(N-k)$
in (\ref{1e}) is just the weak symmetry group $SU(2)_{W}$, as is
traditionally argued \cite{dim}, one is led to $SU(8)$ as the minimal GUT
symmetry ($N-k=2,k/2=3$) \cite{jgk}. Another, and in fact the minimal,
possibility is to identify $SU(N-k)$ with the colour symmetry group $%
SU(3)_{C}$ in the framework of an $SU(7)$ GUT symmetry ($N-k=3,$ $k/2=2$)
\cite{su7tex}. The higher $SU(N)$ GUT solutions, if considered, are also
based on just those two principal possibilities: the weak-component or
colour-component missing VEV vacuum configurations respectively.

Let us see now how this missing VEV mechanism works to solve the
doublet-triplet splitting problem in $SU(8)$ or $SU(7)$ GUT with the
superpotential $W_{A}$ (\ref{2}). It is supposed that there is a
reflection-invariant coupling of the ordinary MSSM Higgs-boson containing
supermultiplets $H$ and $\bar{H}$ with the basic adjoint $\Sigma $, but not
with $\Omega $, in the superpotential $W_{H}$%
\begin{equation}
W_{H}=f_{0}\bar{H}\Sigma H~+W_{H}^{\prime }~~\quad (\Sigma \to -\Sigma 
,\bar{%
H}H\to -\bar{H}H)  \label{4}
\end{equation}
The second part $W_{H}^{\prime }~$contains possible mixings with other
scalar fields, which are inessential for the moment. The superfields $H$ and
$\bar{H}$ do not develop VEVs during the first stage of the symmetry
breaking (\ref{1e}). Thereupon the first term in $W_{H}$ turns into a mass
term for $H$ and $\bar{H}$ determined by the missing VEV pattern (\ref{1d}).
This vacuum, while giving generally heavy masses (of the order of $M_{GUT}$)
to $H$ and $\bar{H}$, leaves their weak components strictly massless. To be
certain of this, we must specify the multiplet structure of $H$ and 
$\bar{H}$
for both the weak-component and the colour-component missing VEV vacuum
configurations, that is in $SU(8)$ and $SU(7)$ GUTs respectively. In the $%
SU(8)$ case $H$ and $\bar{H}$ are fundamental octets whose weak components
(ordinary Higgs doublets) do not get masses from the basic coupling 
(\ref{4}%
). In the $SU(7)$ case $H$ and $\bar{H}$ are 2-index antisymmetric $21$%
-plets in which, after projecting out the extra heavy states (see Section
3.1), there is left just one pair of massless Higgs doublets as a
consequence of the coupling (\ref{4}). Thus, there is a natural
doublet-triplet splitting in both cases and we also have a vanishing $\mu $
term at this stage. However, radiative corrections generate a $\mu $ term of
the right order of magnitude at the next stage when SUSY breaks \cite{jgk}.

\subsection{ Projection to low energies}

Missing VEV vacua, which ensure the survival of the MSSM at low energies,
only appear in $SU(N)$ GUTs with a higher symmetry group than the standard 
$%
SU(5))$ model. In order not to spoil gauge coupling unification, the extra
gauge symmetry should also be broken, $SU(N)\rightarrow SU(5)$, at the GUT
scale. Then the following question arises: how can the missing VEV survive
this extra symmetry breaking with at most a shift of order the electroweak
scale? This requires, in general, that the superpotential (\ref{2}) be
strictly protected from any large influence from the $N-5$ scalars $\varphi
^{k}$ ($k=1,...,N-5$) providing the extra symmetry breaking (or from
uncontrollable gravitational corrections). Technically, such a custodial
symmetry may be a superstring-inherited anomalous $U(1)_{A}$ \cite{gr-sch},
which can naturally keep two sectors of the total superpotential separate
and then induce a high-scale extra symmetry breaking through the
Fayet-Iliopoulos (FI) $D$-term \cite{witten}:

\begin{equation}
D_{A}=\xi +\sum Q_{A}^{k}\mid <\varphi ^{k}>\mid ^{2}, \qquad \xi =\frac{%
TrQ_{A}}{192\pi ^{2}}g_{str}^{2}M_{P}^{2}.  \label{6'}
\end{equation}
Here the sum runs over all ''charged'' scalar fields in the theory,
including those which do not develop VEVs and which contribute to $TrQ_{A}$
only. For realistic or semi-realistic models, $TrQ_{A}$ has turned out to be
quite large, $TrQ_{A}=O(100)$ (see \cite{zz} for a recent discussion).
Therefore, the spontaneous breaking scale of the $U(1)_{A}$ symmetry and of
the related extra gauge symmetry is naturally located at the string scale.
The protecting anomalous $U(1)_{A}$ symmetry is needed to keep the scalars 
$%
\varphi ^{(k)}$ and $\overline{\varphi }^{(k)}$ essentially decoupled from
the basic adjoint superpotential $W_{A}$ (\ref{2}), so as not to strongly
influence its missing VEV vacuum configuration (\ref{1d}). Otherwise
potentially dangerous couplings could appear of the type $\overline{\varphi 
}%
^{(k)}\Sigma \varphi ^{(k)}$, where the $\varphi ^{(k)}$ and $\overline{%
\varphi }^{(k)}$ scalar superfields are taken in pairs of conjugate
fundamental representations ($N$ and $\overline{N}$) of $SU(N)$. If these
couplings actually appeared, they would give rise to shifts in the missing
VEV components of the adjoint scalar $\Sigma _{B}^{A}$ of the order $\Sigma
_{B}^{A}\sim \frac{\delta_{B}^{A}}{m} \overline{\varphi }^{(k)}\varphi
^{(k)}\sim O(M_{GUT})$, as directly follows from the minimisation condition
for the scalar potential. So the presence of a protecting symmetry is
essential for the missing VEV mechanism to function properly.

We will now enlarge on this key point in order to gain a better
understanding of the missing VEV approach. The symmetry protected separation
of the adjoint scalar and the $\varphi ^{k}$ scalar sectors in the total
superpotential implies the appearance of an accidental global symmetry $%
SU(N)_{\Sigma -\Omega }\otimes U(N)\varphi $ in the $SU(N)\otimes $ $%
U(1)_{A} $ gauge theory considered. This global symmetry is in turn
radiatively broken, resulting in a set of pseudo-Goldstone (PG) states of
the type
\begin{equation}
5~+~\bar{5}~+~SU(5)-{\rm singlets}  \label{6}
\end{equation}
which gain a mass at the TeV scale where SUSY softly breaks \cite{jgk}.
There can be a maximum of $N-5$ families of PG states of the type (\ref{6}),
corresponding to the case where the scalars $\varphi ^{(k)}$ and 
$\overline{%
\varphi }^{(k)}$ are only allowed to appear in the Higgs potential through
the basic $SU(N)$ and $U(1)_{A}$ $D$-terms. In this case the $U(N)\varphi $
global symmetry is increased to $U(N)_{\varphi ^{(1)}}\otimes ....\otimes
U(N)_{\varphi ^{(N-5)}}$. This case would occur if the $U(1)_{A}$ charges of
the bilinears $\overline{\varphi }^{k}\varphi ^{k^{\prime }}$ were all
positive (or negative), so that they could not appear in the $SU(N)\otimes
U(1)_{A}$ invariant superpotential in any order.

However, in a properly extended model it is possible for the adjoint and
fundamental scalar sectors in the superpotential to overlap without
disturbing the adjoint missing VEV configuration. This naturally occurs when
the scalars $\varphi ^{(k)}$ are conditioned by the $U(1)_{A}$ symmetry to
develop orthogonal VEVs along the $^{\prime \prime }$extra$^{\prime \prime 
}$
directions
\begin{equation}
\varphi _{A}^{(k)}=\delta _{A,5+k}V_{k},\quad k=1,...,N-5  \label{6''}
\end{equation}
As a result, some safe non-diagonal couplings $\overline{\varphi }%
^{(m)}\Sigma \varphi ^{(n)}$  are generated between the two sectors, 
giving contributions to the pseudo-Goldstone masses which leave only one 
light PG family (\ref{6'}).

Let us consider this possibility in some detail. The least restrictive
choice of such safe mixing terms for the general $SU(N)$ case is achieved by
introducing two sets of new singlet scalar superfields fields, 
$S_{mn}$
and $T_{mn}$, with non-diagonal couplings of the type

\begin{equation}
W_{mix}=\sum_{m<n}^{N-5}\overline{\varphi }^{(m)}[a_{mn}S_{mn}+b_{mn}T_{mn}%
\Sigma ]\varphi ^{(n)}  \label{6'''}
\end{equation}
which are also invariant under the reflection symmetry $\Sigma \rightarrow
-\Sigma $, $T_{mn}\rightarrow -T_{mn}$. The coupling constants $a_{mn}$ and 
$%
b_{mn}$ are all of order $O(1)$ and $O(1/M_{P})$ respectively, and the $%
(N-5)(N-6)/2$ singlet scalars $S_{mn}$ and $T_{mn}$ ($m<n$) get their VEVs
through the FI $D$-term (\ref{6'}), as do all the $\varphi $ and 
$\overline{%
\varphi }$ scalars. One can consider the fields $S_{mn}$ as the basic
carriers of the $U(1)_{A}$ charges $Q_{mn}$ which are all taken positive in
the model (the fields $T_{mn}$ carry the same charges $Q_{mn}$). The $U(1)_{A}$
charges of the $\overline{\varphi }^{(m)}\varphi ^{(n)}$ bilinears ($m<n$)
appearing in $W_{mix}$ are then determined to be $-Q_{mn}$, while the charges 
of all the other bilinears, diagonal $\overline{\varphi }^{(m)}\varphi ^{(m)}$
and non-diagonal $\overline{\varphi }^{(n)}\varphi ^{(m)}$, can always be 
chosen positive. This implies that any terms containing $\varphi $ and 
$\overline{\varphi }$ scalars can only appear in the superpotential
if they also include the bilinears $\overline{\varphi }^{(m)}\varphi
^{(n)}$ so as to properly compensate the $U(1)_{A}$ charges. However, for a
vacuum configuration where the orthogonality conditions $\overline{\varphi 
}%
^{(m)}\varphi ^{(n)}=0$ naturally arise, such terms do not lead (in any
order) to the dangerous $\overline{\varphi }^{(k)}\Sigma \varphi ^{(k)}$
couplings, although they can can contribute to the pseudo-Goldstone
masses. In fact these orthogonality conditions are satisfied at the SUSY
invariant global minimum of the Higgs potential, as follows from the
vanishing $F$-terms of the superfields $S_{mn}$ ($T_{mn}$), $\overline{%
\varphi }^{(m)}$ and $\varphi ^{(n)}$ involved in (\ref{6'''}):

\begin{equation}
\overline{\varphi }^{(m)}\varphi ^{(n)}=0,\quad \quad
a_{mn}S_{mn}=-b_{mn}T_{mn}\Sigma _{5+n}^{5+n},\quad \quad m<n
\label{7a}
\end{equation}
(no summation is implied). Here the orthogonal VEV values of the scalars $%
\varphi ^{(n)}$ (\ref{6''}) have been used. One can now readily see that
non-diagonal mass terms appear for the PG states related to the multiplets 
$%
\overline{\varphi }^{(m)}$ and $\varphi ^{(n)}$

\begin{equation}
M_{mn}\equiv [W_{mix}^{^{\prime \prime }}]_{\overline{\varphi }^{(m)}\varphi
^{(n)}}=b_{mn}T_{mn}(\Sigma -\Sigma _{5+n}^{5+n}\cdot I),\quad \quad m<n
\label{7b}
\end{equation}
where $I$ is the $N\times N$ unit matrix. Diagonalisation of the mass matrix
(\ref{7b}) explicitly shows that one PG superposition $5~+~\bar{5}$ 
(\ref{6}%
) is left massless, while the others become heavy\footnote{%
This mass matrix is in fact a ''triangular'' $(N-5)\times (N-5)$ matrix with
zeros on the main diagonal, $M_{mn}=0$ for $m\geq n$. Such a matrix has in
general only one zero eigenvalue.}. This is in fact a general consequence of
the symmetry breaking pattern involved. The point is that neither of the
other mass-terms $M_{mm}$ and $M_{nm}$ can be allowed by $U(1)_{A}$ symmetry
for any generalisation of the superpotential $W_{mix}$ (\ref{6'''}).
Otherwise the dangerous $\overline{\varphi }^{(k)}\Sigma \varphi ^{(k)}$
couplings inevitably appear as well. So, one can conclude that even in the
general case one PG family of the type (\ref{6}) always exists. Together
with the ordinary quarks and leptons and their superpartners these PG
states, both bosons and fermions, determine the particle spectrum at low
energies. In most of what follows the existence of just one family of PG
states at the sub-TeV scale will be assumed.

We consider below both of the minimal possible GUTs, $SU(7)$ and $SU(8)$,
with the missing VEV solution naturally allowing the survival of the MSSM
down to low energies. Whereas the $SU(7)$ model is taken as an ordinary
one-family unifying GUT \cite{su7tex}, the $SU(8)$ model can include
unification of the quark-lepton families as well \cite{su8}.

\section{One-family unifying GUT: $SU(7)$}

By analogy with the standard $SU(5)$ model, we take the simplest
anomaly--free set of matter fields, consisting of the combination of the
fundamental and 2-index antisymmetrical representations of the $SU(7)$ gauge
group

\begin{equation}
\left[ 2\overline{\Upsilon }^{A}+\overline{\Psi }^{A}+\Psi _{[AB]}\right]
_{i}  \label{7}
\end{equation}
($A,B=1,...,7$ are the $SU(7)$ indices) for each of the three quark--lepton
families or generations ($i=1,2,3)$. The quarks and leptons belong to the
multiplets $\overline{\Psi }^{A}(\overline{7})+$ $\Psi _{[AB]}(21)$ , while
the extra multiplets $\overline{\Upsilon }^{A}$ are specially introduced in 
(%
\ref{7}) for anomaly cancellation.

There is also a set of Higgs superfields among which are the two already
mentioned adjoint Higgs multiplets $\Sigma _{B}^{A}$ and $\Omega _{B}^{A}$,
responsible for the breaking (\ref{1e}, \ref{3b}) of $SU(7)$, and a
conjugate pair of multiplets $H_{[AB]}$ and $\bar{H}^{[AB]}$ (being the 
$21$%
-plets of the $SU(7)$) where the ordinary electroweak doublets $H_{u}$ and 
$%
\overline{H}_{d}$ reside. Besides, as in the general $SU(N)$ case (see
Section 2.2), there should be extra-symmetry breaking scalar superfields $%
\varphi ^{(p)}$ and $\overline{\varphi }^{(p)}$ ($p=1,2$) which are taken to
be fundamental septets and anti-septets respectively. They are supposed to
develop their string-scale order VEVs along the ''extra'' directions
\begin{equation}
\varphi _{A}^{(1)}=\delta _{A6}V_{1,} \quad \varphi _{A}^{(2)}=\delta
_{A7}V_{2}  \label{17}
\end{equation}
only through the (FI) $D$-term related to the $U(1)_{A}$ symmetry (\ref{6'}%
). The protecting anomalous $U(1)_{A}$ symmetry keeps the $\varphi $ scalars
decoupled from the basic adjoint superpotential $W_{A}$ (\ref{2}), so as not
to strongly influence the missing VEV solution (\ref{1d}) through dangerous
couplings of the type $\overline{\varphi }^{(p)}\Sigma \varphi ^{(p)}$.

With the given assignment of matter and Higgs superfields, the particle
spectrum at low energies looks as if one had just the standard SUSY $SU(5)$
as a starting GUT symmetry, except that one family of PG states of type 
(\ref
{6}) appears, when a missing VEV vacuum configuration develops in the 
$SU(7)$
GUT. With this exception, all the other $SU(7)$ inherited states in matter
and Higgs multiplets aquire GUT scale masses due to symmetry breaking, thus
completely decoupling from low-energy physics. We demonstrate this for the
Higgs sector in the next sub-section.

\subsection{Higgs sector}

We now show that all the states, except for one pair of weak doublets in the
basic Higgs multiplets $H_{[AB]}$ and $\overline{H}^{[AB]}$, become
superheavy. Firstly one substitutes the colour-component missing VEV
solution, obtained from the general case (\ref{1d}) by setting $N=7$ and $%
k=4 $, into the superpotential (\ref{4}). Superheavy masses are thereby
generated for most of the components of the $H$ and $\overline{H}$
multiplets. However, the following states (weak, colour and extra symmetry
components are explicitly indicated)

\begin{equation}
H_{w6}, \quad \overline{H}^{w6}, \quad H_{w7}, \quad \overline{H}^{w7},
\quad H_{[cc^{\prime }]}, \quad \overline{H}^{[cc^{\prime }]}  \label{23a}
\end{equation}
still remain massless at this stage of $SU(7)\ $symmetry breaking (\ref{1e}%
). Therefore one of the two pairs of weak doublets in (\ref{23a}), as well
as the colour triplets, must further become heavy in order to get the
ordinary picture of MSSM at low energies. This happens as a result of mixing
$H$ and $\overline{H}$ with the specially introduced heavy scalar
supermultiplets $\Phi _{[ABC]}$ and $\overline{\Phi }^{[ABC]}$ (being $35$%
-plets of $SU(7)$) in the basic Higgs superpotential

\begin{equation}
W_{H}^{\prime }=f\cdot H\overline{\Phi }\varphi ^{(1)}+\overline{f}\cdot
\overline{H}\Phi \overline{\varphi }^{(1)}+y\cdot S\overline{\Phi }\Phi ,
\label{www}
\end{equation}
($f$, $\overline{f}$ and $y$ are dimensionless coupling constants) when the
scalars $\varphi $ get their VEVs, thus breaking the extra gauge symmetry.
The presence of the $^{\prime \prime }$conjugated$^{\prime \prime }$ $%
\overline{\Phi }-H\ $ and $\Phi -\overline{H}$ mixings in $W_{H}^{\prime }$
could allow the dangerous $\overline{\varphi }\Sigma \varphi $ terms,
destroying the missing VEV solution, unless the bilinear term $\overline{%
\Phi }\Phi $ has nonzero $U(1)_{A}$ charge. Therefore, this term appears in 
$%
W_{H}^{\prime }$ together with the singlet scalar superfield $S$, the basic 
$%
U(1)_{A}$ charge carrier introduced earlier in $W_{mix}$ (\ref{6'''}) in a
general $SU(N)$ context (for $SU(7)$ there appears only one pair of such
singlets, $S$ and $T$).

It should be clear now that the $W_{H}^{\prime }$ couplings (\ref{www}) will
rearrange the mass spectrum of the states (\ref{23a}), so as to leave just
one pair of massless weak-doublets as needed for the MSSM. By diagonalising
the $2\times 2$ mass matrix for the states $H_{[cc^{\prime }]}$ and $%
\overline{H}^{[cc^{\prime }]}$ and the double-coloured components $%
\Phi_{[cc^{\prime }6]}$ and $\overline{\Phi }^{[cc^{\prime }6]}$, the mass
of the colour triplet components in (\ref{23a}) is found to be of order
\begin{equation}
M_{*}\sim \frac{f\overline{f}}{y}\frac{<\varphi ><\overline{\varphi }>}{S}%
\sim M_{GUT}  \label{23b}
\end{equation}
where the combination of the primary coupling constants $f,$ $\overline{f}$%
and $y$, can be taken $O(1)$ in general. There is a $3\times 3$ mass matrix
for the weak doublet states, corresponding to the mixing of the states $%
H_{w6}$, $H_{w7}$ and $\Phi _{[w67]}$ and their $^{\prime \prime } $%
conjugates$^{\prime \prime }$ $\overline{H}^{w6}$, $\overline{H}^{w7}$ and 
$%
\overline{\Phi }^{[w67]}$ respectively. After diagonalisation this matrix
leaves just one pair of weak-doublets $H^{w6}$ and $\overline{H}^{w6}$
strictly massless, while the other pair $H_{w7}$ and $\overline{H}^{w7}$
aquires a mass $M_{*}$ (\ref{23b}) of order $M_{GUT}$.

In much the same way all the additional states in the $SU(7)$ matter
multiplets (\ref{7}) become superheavy during the starting GUT symmetry
breaking $SU(7)\rightarrow SU(5)$ \cite{su7tex}.

\subsection{Yukawa couplings}

The usual dimension-4 trilinear Yukawa couplings are forbidden by $SU(7)$
gauge invariance. So we suppose that all the generalized Yukawa couplings,
the $RP$-conserving (ordinary up and down fermion Yukawas) as well as the $%
RP $-violating ones allowed by the $SU(7)\otimes U(1)_{A}$ symmetry, are
given by a similar set of dimension-5 operators of the form ($i,j,k=1,2,3$
are the generation indices, the $SU(7)$ indices $A,B,C=1,...,7$ are
hereafter omitted):
\begin{equation}
{\cal O}_{ij}^{up}=\frac{G_{ij}^{u}}{M_{P}}(\Psi _{i}\Psi _{j})(H\varphi
^{(2)})  \label{14}
\end{equation}
\begin{equation}
{\cal O}_{ij}^{down}=\frac{G_{ij}^{d}}{M_{P}}(\overline{\Psi }_{i}\Psi 
_{j})(%
\overline{H}\varphi ^{(1)})  \label{15}
\end{equation}
\begin{equation}
{\cal O}_{ijk}^{rpv}=\frac{G_{ijk}}{M_{P}}(\overline{\Psi }_{i}\Psi _{j})(%
\overline{\Psi }_{k}\Sigma ).  \label{16}
\end{equation}

Further, substituting the VEVs of the scalars $\Sigma $ (\ref{1d}) and $%
\varphi $ (\ref{17}) into the basic operators (\ref{14}--\ref{16}), one
obtains at low energies the effective renormalisable Yukawa and LNV
interactions with coupling constants

\begin{equation}
Y_{ij}^{u}=G_{ij}^{u}\frac{<\varphi ^{(2)}>}{M_{P}} , \quad
Y_{ij}^{d}=G_{ij}^{d}\frac{<\varphi ^{(1)}>}{M_{P}} , \quad \Lambda
_{ijk}=G_{ijk}\frac{<\Sigma >}{M_{P}} .  \label{Y}
\end{equation}
At the same time the baryon number non-conserving couplings $%
\lambda_{ijk}^{\prime \prime}$ completely disappear. The crucial point is
that the adjoint field $\Sigma $ develops a VEV configuration with strictly
zero colour components (\ref{1d}) in the SUSY limit. When SUSY breaks,
radiative corrections will shift the missing VEV components of $\Sigma $ to
nonzero values of order $M_{SUSY}$, thus inducing the ordinary $\mu $-term
of the MSSM, on the one hand, and baryon number violating interactions with
hierarchically small coupling constants of the order $M_{SUSY}/M_{GUT}$, on
the other.

The effective dimension-5 interactions (\ref{14}--\ref{16}) could be
generated by the exchange of some heavy states, such as massive string
modes. When generated by the exchange of the same superheavy multiplet (that
is a vector-like pair of fundamental septets $7+\overline{7}$), the
resulting operators (\ref{15}) and (\ref{16}) have effective coupling
constants (\ref{Y}) aligned in flavour space \cite{JClnv}:

\begin{equation}
\Lambda _{ijk}=Y_{ij}^{d}\cdot \epsilon _{k}^{\ }  \label{22a}
\end{equation}
The parameters $\epsilon _{k}^{\ }$ ($k=1,2,3)$ include some known
combination of the primary dimensionless coupling constants and a ratio of
the VEVs of the scalars $\Sigma$ and $\varphi$. This relation (\ref{22a})
further splits into the ones for charged lepton ($cl$) and down quark ($dq)$
LNV couplings respectively,

\begin{equation}
\lambda _{ijk}=Y_{ij}^{cl}\cdot \epsilon _{k} ,\qquad \lambda _{ijk}^{\prime
}=Y_{ij}^{dq}\cdot \epsilon _{k} ,
\end{equation}
when evolved from the $SU(7)$ scale down to low energies.

So we see that the possible common origin of all the generalised Yukawa
couplings, both $RP$-conserving and $RP$- violating, at the GUT scale
results in some minimal form of lepton number violation, provided that the
appropriate heavy-state mediator exists. As a result, we are driven to a
simple picture where the flavour structure, as well as the hierarchies of
the trilinear LNV couplings in $\Delta W$ (\ref{1}), are essentially aligned
with the down quark and charged lepton mass and mixing hierarchies. At the
same time, the effective bilinear LNV terms appear to be generically
suppressed by the custodial $U(1)_{A}$ symmetry (for a detailed exposition
see a recent paper \cite{JClnv}).

At low energies, the minimal LNV model presented can be viewed as an
alternative to another minimal model based on the MSSM, in which only the
bilinear LNV terms $\mu _{i}L_{i}H_{u}$ in $\Delta W$ (\ref{1}) are included
\cite{lnvGUTs, Dudas}. Depending on the $U(1)_{A}$ charges assigned to the
matter and Higgs superfields involved, one can generically obtain at low
energies either the bilinear model or the trilinear one considered here. The
bilinear model also leads to LNV-Yukawa coupling alignment, by virtue of
which many predictions of both models concerning quark flavour conservation
are very similar \cite{JClnv}. However, there are principal differences as
well. The point is that the influence of the SUSY soft breaking sector,
being predominant for the bilinear model, is quite negligible for the
present one. Therefore, the LNV-Yukawa alignment, while appearing in both
models, leads in the latter case to distinctive flavour-dependent relations
between various LNV processes arising from slepton and squark exchanges
(which are basically conditioned by the quark and lepton mass hierarchy)
\cite{JClnv}. By contrast, in the bilinear model these processes appear to
be essentially determined by $W$ and $\ Z$ bozon exchanges and, as a result,
are largely flavour-independent. On the other hand, the bilinear model has a
serious problem of extension to the GUT framework. Any such extension leads,
together with a lepton mixing with a weak Higgs doublet, to a quark mixing
with a colour Higgs triplet, thus inducing baryon number violation as well.
The only handle one has to address this problem seems to be the use of
electroweak scale masses $\mu _{i}$ in the GUT-symmetry invariant bilinear
couplings. Their use would mean that new fine-tuning conditions, besides the
ordinary gauge hierarchy one, should be satisfied in a very ad hoc way.

An extended discussion of the properties of the $SU(7)$ GUT, including the
solution to the doublet-triplet splitting problem, string scale unification,
proton decay, hierarchy of baryon vs lepton number violation and neutrino
masses, can be found in our recent paper \cite{su7tex}.

\section{Three-family unifying GUT: $SU(8)$}

It is tempting to treat the extra gauge symmetry in a general $SU(N)$ GUT as
a flavour symmetry. If so, according to the particular solution ({\ref{1d}) 
}
for the weak-component missing VEV configuration, the numbers of fundamental
colours and flavours must be equal ($n_{C}=n_{F}=k/2$) for any even-order $%
SU(N)$ group, among which the minimal one is $SU(8)$ ($n_{C}=n_{F}=3$).
Thus, in the $SU(8)$ case, the missing VEV configuration requires an
additional colour-flavour symmetry: $SU(3)_{C}$ $\leftrightarrow $ $%
SU(3)_{F} $.

Having considered the basic matter superfields (quarks and leptons and their
superpartners), the question of whether the above flavour symmetry $%
SU(3)_{F} $ is really their family symmetry naturally arises. Needless to
say, among many other possibilities, the special assignment treating the
families as a fundamental triplet of $SU(3)_{F}$ is the most attractive. In
such a case the anomaly-free set of $SU(8)$ antisymmetric multiplets (in a
self-evident notation; $A,B,C=1,2,....,8$ )

\begin{equation}
6\cdot \overline{8}^{A}+\overline{28}^{[AB]}+2\cdot 56_{[ABC]}+70_{[ABCD]}
\label{25}
\end{equation}
is singled out, if we require that after flavour symmetry breaking only
three massless families of ordinary quarks and leptons (and their
superpartners) are left as chiral triplets of $SU(3)_{F}$, stemming from the
multiplets

\begin{equation}
\overline{28}=(\bar{5},\bar{3})+...,~~~~70=(10,\bar{3})+...  \label{26}
\end{equation}
The remaining $SU(5)\otimes SU(3)_{F}$ components, in these as well as in
the other multiplets (\ref{25}), acquire heavy masses of order $M_{F}\sim
M_{GUT}$ \footnote{%
The special multiplet arrangement (\ref{25}) was considered before by one of
us \cite{su8} as a possible basis for the family-unifying $SU(8)$ GUT.
Remarkably, the multiplets (\ref{25}) follow from the unique (''each
multiplet - one time'') set of $SU(11)$ multiplets \cite{su11} after the
symmetry breaking $SU(11)\to SU(8)$ and the exclusion of all the conjugated
(under $SU(8)$) multiplets except the self-conjugated one $70_{[ABCD]}$.}.
So, one arrives at the chiral $SU(3)_{F}$ family symmetry case \cite{su3H},
which leads to a natural conservation of flavour both in the particle and
sparticle sectors.

Furthermore, there is a universal see-saw mechanism in the $SU(8)$ model,
with heavy intermediate states provided by the multiplets (\ref{25}), which
induces non-trivial fermion mass-matrices with many texture ans\"{a}tze
available. So the observed pattern of quark and lepton masses and mixings
can appear once the electroweak $SU(2)\otimes U(1)_{Y}$ symmetry breaks 
\cite
{su8}. At the same time, by analogy with the $SU(7)$ case, see (\ref{16}),
the only $RP$-violating coupling allowed by $SU(8)\otimes U(1)_{A}$ symmetry
is supposed to be given by the dimension-5 operator

\begin{equation}
{\cal O}_{rpv}\propto \frac{1}{M_{P}}(\overline{\Psi }^{[AB]\ }\Psi 
_{[ABCD]}%
\overline{\Psi }^{[CD^{\prime }]}\Sigma _{D^{\prime }}^{D})  \label{27}
\end{equation}
Here the matter fields $\overline{\Psi }$ and $\Psi $ belong to the basic
multiplets{\ (\ref{26}). One can see now that the weak-component missing VEV
solution for }$\Sigma $ (\ref{1d}), when substituted into the operator $%
{\cal O}_{rpv}$, leaves only the LNV couplings and projects out the baryon
number violating ones. At low energies the surviving effective couplings
take the form
\begin{equation}
\lambda \epsilon _{\alpha \beta \gamma }{\cal (}L^{\alpha }L^{\beta }%
\overline{E}^{\gamma }+rL^{\alpha }Q^{\beta }\overline{D}^{\gamma })
\label{28}
\end{equation}
Here $\alpha ,$ $\beta ,$ $\gamma =1,2,3$ are the generation indices,
belonging to the family $SU(3)_{F}$ symmetry, and $r$ is a factor giving the
relative coupling constant renormalisation after evolution from the $SU(8)$
scale to low energies. So, as in the $SU(7)$ case, one has baryon number
conservation at the same time as lepton number violation in the SUSY limit.

Meanwhile, despite their common origin, there is a principal difference
between the $SU(7)$ and $SU(8)$ cases.{\ The point is that the basic adjoint
$\Sigma $ moduli mass ratio $M_{3}/M_{8}$ appears, according to the missing
VEV vacua (\ref{1d}), to be 2 and 1/2 for $SU(7)$ and $SU(8)$ respectively.
As was shown in recent papers \cite{su7tex, su5tex}, this ratio essentially
determines the high-energy behavior of the MSSM gauge couplings. In fact it
follows that the unification scale in $SU(7)$ is pushed to the string scale
\cite{su7tex}, while the unification scale in $SU(8)$ ranges, at best, close
to the standard unification value \cite{jgk}.}

\section{Conclusions}

The absence of automatic global conservation laws in SUSY theories, in
contrast to the Standard Model, is frequently considered as a drawback of
supersymmetry. Meanwhile phenomenologically, whereas SUSY-inspired B number
non-conservation must be highly suppressed, SUSY-inspired L-number violation
could occur at a level large enough for the observation of its many
spectacular manifestations \cite{lnv,JClnv}. One of these manifestations may
be the sizeable atmospheric neutrino oscillations recently reported 
\cite{SK}%
, according to which one of the neutrino species is expected to have a mass
at least of order $0.1$ eV. That means, in general, the particle content of
the MSSM or the minimal $SU(5)$ SUSY GUT should be extended to include new
states, that is fundamentally heavy right-handed neutrinos or even light
sterile left-handed ones. Neutrino masses per se do not yet give any
conclusive evidence in favour of SUSY theories. However sizeable LNV in the
charged lepton sector and, of course, in the decays of the lightest
supersuymmetric particle \cite{JClnv}, if actually observed, could qualify
as generic SUSY inspired phenomena. In such a situation the following
question would arise, which should be addressed within the framework of
Grand Unification rather than the MSSM: what could stand behind such a
tremendous hierarchy of lepton vs baryon number violation?

In this connection we suggested that the nature of the global conservation
laws in SUSY theories is determined by the basic vacuum configuration which
breaks the underlying GUT symmetry. Following this idea, we have argued that
the GUTs with a natural missing VEV solution to the doublet-triplet
splitting problem could, simultaneously, provide the reason for treating
lepton and baryon number carrying matter fields differently. We have shown
that missing VEV vacuum configurations, ensuring the survival of the MSSM
gauge symmetry at low energies, only emerge in extended SU(N) GUTs with $%
N\ge 7$. Further, the one-family unifying $SU(7)$ and the three-family
unifying $SU(8)$ GUTs have been constructed. In both cases the effective LNV
couplings immediately evolve from the GUT scale, while the baryon number
non-conserving ones are safely projected out by the missing VEV vacuum
configuration, which breaks the starting GUT symmetry down to that of the
MSSM. However, at the next stage when SUSY breaks, radiative corrections
shift the missing VEV to some nonzero value of order $M_{SUSY}$, thus
inducing the ordinary $\mu $-term of the MSSM, on the one hand, and BNV
couplings with the hierarchically small constants $\lambda _{ijk}^{\prime
\prime }=O(M_{SUSY}/M_{GUT})$, on the other. So, a missing VEV solution to
the gauge hierarchy problem leads, in a literal sense, to the same hierarchy
of baryon vs lepton number violation.

\section*{Acknowledgments}

We would like to thank many of our colleagues, especially Riccardo Barbieri,
Grahame Blair, Ilia Gogoladze, Mike Green, David Hutchcroft, Archil
Kobakhidze, Gordon Moorhouse, Alexei Smirnov and David Sutherland for
stimulating dicussions and useful remarks. Financial support by the INTAS
grants No. RFBR 95-567, 96-155, PPARC grant No. PPA/V/S/1997/00644 and the
Joint Project grant from the Royal Society are also gratefully acknowledged.


\begin{thebibliography}{99}
\bibitem{nilles}  H.P. Nilles, Phys.Rep. 110 (1984) 1.

\bibitem{2}  S. Dimopoulos and H. Georgi, Nucl.Phys. B 193 (1981) 150;

N. Sakai, Z.Phys. C 11 (1981) 153;

S. Weinberg, Phys. Rev. D 26 (1982) 287;

N. Sakai and T. Yanagida, Nucl. Phys. B 197 (1982) 533.

\bibitem{lnv}  H.K. Dreiner, in: {\it Perspectives in Supersymmetry}, ed. by
G.L. Kane, World Scientific, hep--ph/9707435;

G. Bhattacharyya, hep--ph/9709395;

R. Barbier et al., hep--ph/9810232.

\bibitem{Z3}  L.E. Iba\~{n}ez and G.G. Ross, Phys. Lett. B 260 (1991) 291;
Nucl. Phys. B 368 (1992) 3.

\bibitem{lnvGUTs}  L.J. Hall and M. Suzuki, Nucl. Phys. B 231 (1984) 419;

D. Brahm and L. Hall, Phys. Rev. D 40 (1989) 2449;

K. Tamvakis, Phys. Lett. B 382 (1996) 251;

R. Hempfling, Nucl. Phys. B 478 (1996) 3;

A.Yu. Smirnov and F. Vissani, Nucl. Phys. B 460 (1996) 37;

G. Giudice and R. Ratazzi, Phys. Lett. B 406 (1997) 321;

R. Barbieri, A. Strumia and Z. Berezhiani, Phys. Lett. B 407 (1997) 250 .

\bibitem{jgk}  J. Chkareuli, Talk given at {\it Trieste Conference on Quarks
and Leptons: Masses and Mixings}, Trieste, 1996;{\it \ }hep-ph/9706280;

J.L. Chkareuli and A.B. Kobakhidze, Phys. Lett. B 407 (1997) 234;

J.L. Chkareuli, I.G. Gogoladze and A.B. Kobakhidze, Phys. Rev. Lett. 80
(1998) 912;

J.L. Chkareuli, Proceedings of 29th International Conference on High Energy
Physics, Vancouver, 1998, p1669, vol 2 (Eds A. Astbury, D. Axen and J.
Robinson), World Scientific, 1999; hep-ph/9809464.

\bibitem{dim}  S. Dimopoulos and F. Wilczek, in: Erice Summer Lectures
(Plenum, New-York, 1981).

\bibitem{yan}  C. Bachas, C. Fabre and T. Yanagida, Phys. Lett. B 370 (1996)
49;

M. Bastero-Gil and B. Brahmachari, Phys. Lett. B403 (1997) 51.

\bibitem{su7tex}  J.L. Chkareuli, C.D. Froggatt, I.G. Gogoladze and A.B.
Kobakhidze, {\it From Prototype }$SU(5)${\it \ to Realistic }$SU(7)${\it \
SUSY GUT}, hep-ph/0003007.

\bibitem{gr-sch}  M. Green and J. Schwarz, Phys. Lett. B 149 (1998) 117.

\bibitem{witten}  M. Dine, N. Seiberg and E. Witten, Nucl. Phys. B 289
(1987) 587.

\bibitem{zz}  Z. Berezhiani and Z. Tavartkiladze, Phys. Lett. B 396 (1997)
150.

\bibitem{su8}  J.L. Chkareuli, Phys. Lett. B 300 (1993) 361.

\bibitem{su5tex}  J.L. Chkareuli and I.G. Gogoladze, Phys. Rev. D 58 (1998)
055011.

\bibitem{JClnv}  J.L. Chkareuli, I.G. Gogoladze, M.G.\ Green and D.E.
Hutchcroft, A.B. Kobakhidze, {\it On SUSY inspired minimal lepton number
violation}, \- hep-ph/9908451, Phys. Rev. D (to appear).

\bibitem{Dudas}  F. de Campos, M.A. Garcia-Jare\~{n}o, A.S. Joshipura, J.
Rosiek and J.W.F. Valle, Nucl. Phys. B 451, 3 (1995) ;

T. Banks, T. Grossman, E. Nardi and Y. Nir, Phys. Rev. D 52, 5319 (1996);

H.P. Nilles and N. Polonsky, Nucl. Phys. B 484, 33 (1997);

P. Binetruy, E. Dudas, S. Lavignac and C.A. Savoy, Phys. Lett. B 422 (1998)
171.

\bibitem{su11}  H. Georgi, Nucl. Phys. B 156 (1979) 126.

\bibitem{su3H}  J.L. Chkareuli, JETP Lett. 32 (1980) 671;

Z.G. Berezhiani and J.L. Chkareuli, Yad. Fiz. 37 (1983) 1043;

F. Wilczek, preprint NSF-ITP-83-08 (1983);

Z.G. Berezhiani, Phys. Lett. B129 (1983) 99; ibid B150 (1985) 117;

J.L. Chkareuli, JETP Lett. 41 (1985) 577;

J.C. Wu, Phys. Rev. D36 (1987) 1514;

J.L. Chkareuli, JETP Lett. 50 (1989) 255; Phys. Lett. B246 (1990) 498;

T.M. Bibilashvili and G.R. Dvali, Phys. Lett. B248 (1990) 259;

J.L. Chkareuli, JETP Lett. 54 (1991) 189; ibid 54 (1991) 295; Phys. Lett.
B272 (1991) 207;

Z. Berezhiani, Phys. Lett. B 417 (1998) 287.

\bibitem{SK}  The Super-Kamiokande Collaboration (Y.Fukuda et al.),
Phys.Rev.Lett${\em .}$ 81 (1998) 1562 .

\end{thebibliography}
\end{document}